\documentclass[12pt]{article}
\usepackage{graphicx}
\newcommand{\bi}{\bibitem}
\newcommand{\cent}{\centerline}

\newcommand{\Vbf}{\mbox{\boldmath $V$}}
\newcommand{\ubf}{\mbox{\boldmath $u$}}
\newcommand{\imp}{\mbox{\boldmath $p$}}

\newcommand{\kbf}{\mbox{\boldmath $k$}}

\newcommand{\Le}{{\cal L}}

\newcommand{\crm}{{\rm c}}
\newcommand{\drm}{{\rm d}}
\newcommand{\erm}{{\rm e}}
\newcommand{\Irm}{{\rm I}}
\newcommand{\IIrm}{{\rm II}}
\newcommand{\IIIrm}{{\rm III}}

\newcommand{\grm}{{\rm g}}

\newcommand{\wrm}{{\rm w}}
\newcommand{\inrm}{{\rm in}}
\newcommand{\ga}{\gamma}

\newcommand{\infi}{\infty}

\newcommand{\lan}{\langle}

\newcommand{\De}{\Delta}
\newcommand{\ran}{\rangle}
\newcommand{\bb}{\begin{equation}}
\newcommand{\ee}{\end{equation}}
\newcommand{\bega}{\begin{eqnarray}}
\newcommand{\ega}{\end{eqnarray}}
\newcommand{\begae}{\begin{eqnarray*}}
\newcommand{\egae}{\end{eqnarray*}}

\newcommand{\up}{\uparrow}
\newcommand{\dow}{\downarrow}

\newcommand{\h}{\hspace*{4ex}}
\newcommand{\disp}{\displaystyle}

\newcommand{\longr}{\longrightarrow}

\newcommand{\ove}{\overline}

\newcommand{\vs}{\vspace*}

\begin{document}

\baselineskip 0.6cm

\cent{\large{{\bf About Superluminal motions and Special
Relativity:}}} \cent{\large{{\bf A Discussion of some recent
Experiments,}}} \cent{\large{{\bf and the solution of the Causal
Paradoxes}} \footnote{Work partially supported by INFN, MURST and
CNR, and by CAPES. \ E-mail address for contacts:
Recami@mi.infn.it }}

\

\cent{Erasmo Recami$^{(a)}$, \ Flavio Fontana$^{(b)}$ \ {\rm and} \
Roberto Garavaglia$^{(c)}$}

\

\cent{$^{(a)}$ {\em Facolt\`a di Ingegneria, Universit\`a Statale
di Bergamo, Bergamo, Italy;}} \cent{{\rm and} {\em INFN--Sezione
di Milano, Milan, Italy; }} \cent{$^{(b)}$ {\em R \& D division,
Pirelli Labs, Milan, Italy.}} \cent{$^{(c)}$ {\em P.R.B. s.r.l.,
Milan, Italy;} {\rm and}} \cent{{\em Dip.to di Scienze
dell'Informazione, Universit\`a di Milano, Milan, Italy.}}

\begin{abstract}
Some experiments, performed at Berkeley, Cologne, Florence,
Vienna, Orsay, Rennes, etc., led to the claim that something seems
to travel with a group velocity larger than the speed $c$ of light
in vacuum. Various other experimental results seem to point in the
same direction: For instance, localized wavelet-type solutions of
Maxwell equations have been found, both theoretically and
experimentally, that travel with Superluminal speed. Even muonic
and electronic neutrinos  ---it has been proposed---  might be
``tachyons", since their square mass appears to be negative. \
With regard to the first-mentioned experiments, it was recently
recently claimed by Guenter Nimtz that those results with
evanescent waves (or tunneling photons)  imply superluminal signal
and impulse transmission, and therefore violate Einstein
causality. \ In this note, on the contrary, we want to stress that
all such results do {\em not} place relativistic causality in
jeopardy, even if they refer to actual tachyonic motions: In fact,
Special Relativity can cope even with Superluminal objects and
waves. \ For instance, it is possible (at least in microphysics)
to solve also the known causal paradoxes, devised for ``faster
than light" motion, although this is not widely recognized yet. \
Here we show, in detail and rigorously, how to solve the oldest
causal paradox, originally proposed by Tolman, which is the kernel
of many further tachyon paradoxes (like J.Bell's, F.A.E.Pirani's,
J.D.Edmonds' and others').  {\em The key to the solution is a
careful application of tachyon mechanics, as it unambiguously
follows from special relativity}. \ At Last, in one of the two
Appendices, we propose how to evaluate the group-velocity in the
case of evanescent waves.
\end{abstract}

\


\newpage

\section*{1. Introduction}
Superluminal propagation seems to have been observed in several areas of
physics\cite{1}. \
In electromagnetism, in particular, some recent experiments performed at
Berkeley\cite{2}, Cologne\cite{3}, Florence, Vienna, Orsay and Rennes with
evanescent waves (or ``tunnelling photons") led to the claim that evanescent
modes can travel with a group velocity larger than the speed
$c$ of light in vacuum, thus confirming some older predictions\cite{4}. \
Even more recently, some of the main experimental claims (e.g., in \cite{3,2})
have been shown to be theoretically sound just by solving the (classical)
Maxwell equations with the requested boundary conditions\cite{5}, or by
analysing the corresponding (quantum) tunnelling problems\cite{5}.

\h Various other
experimental results seem to point in the same direction: For instance,
localized wavelet-type solutions of Maxwell equations have been found, both
theoretically\cite{1,5} and experimentally\cite{6}, that travel with
Superluminal speed. \ Even muonic and electronic neutrinos  ---it has been
proposed--- might be ``tachyons", since their square mass appears to be
negative\cite{7}. \ As to the {\em apparent} Superluminal expansions
observed in the core of quasars\cite{8} and, recently, in the so-called
galactic microquasars\cite{9}, we shall not deal here with that problem,
too far from the topics of this paper: without mentioning that the
astrophysical data are often the most difficult to be interpreted.

\h Let us confine our attention to the first-mentioned
experiments\cite{2,3,10}. \ From a
historical point of view, let us recall that for long time the topic of the
electromagnetic wave propagation velocity was regarded as already settled
down by the classical works of Sommerfeld\cite{11} and Brillouin\cite{12}. \
A few authors, however, studying the propagation of light pulses in anomalous
dispersion (absorbing) media both theoretically\cite{13} and experimentally\cite{14},
had found their envelope speed to be the group velocity $v_\grm$, even when
$v_\grm$ exceeds $c$, equals $\pm \infty$, or becomes negative! \ In the
meantime, evanescent waves were predicted\cite{15} to be faster-than-light just
on the basis of Special Relativistic considerations.

\h But evanescent waves in suitable (``undersized") waveguides,
in particular, can be regarded also as tunnelling photons\cite{16}, due to the
known formal analogy\cite{17} between the Schroedinger equation in presence of
a potential barrier and the Helmholtz equation for a wave-guided beam. And
it was known since long that tunnelling particles (wave packets) can move
with Superluminal group velocities inside opaque barriers\cite{18}; therefore,
even from the quantum theoretical point of view, it was expected\cite{18,16,15}
that evanescent waves could be Superluminal.

\h The point we are more interested in is just
the propagation in waveguides of pulses obtained by amplitude modulation
of a carrier-wave endowed with an under-cutoff frequency; \ and
the experiments ---for instance--- in refs.\cite{2,3,19,20} seem to have detected
in such a case a Superluminal group-velocity, $v_\grm > c \;$, \ in agreement
with the classical\cite{15} and the quantum\cite{18} predictions). \  For
example,
the work in refs.\cite{3,21} put in particular evidence the fact that the
segment of ``undersized" (= operating with under-cutoff frequencies) waveguide
provokes an attenuation of each spectral component, without any phase
variation. More precisely, the unique phase variation detectable is due to
the discontinuities in the waveguide cross-section (cf. also refs.\cite{18}
and \cite{5}). \ As stressed, e.g., by Barbero et al.\cite{5}, the spectrum
leaving an undersized waveguide segment
(or {\em photonic barrier\/}) is simply the entering spectrum multiplied by
the transfer function \ $H(\omega) {\; =
\;} \exp[i\beta L]$, \ with \ $\beta(\omega) {\; = \;} \omega \sqrt{
1-(\omega_\crm/\omega)^2}/c$. \ For $\omega > \omega_\crm$, the propagation
constant $\beta(\omega)$ is real, and $H(\omega)$ represents a phase
variation to be added to the outgoing spectrum. \ However, for $\omega <
\omega_\crm$, when $\beta(\omega)$ is imaginary, the transfer function
just represents an additional attenuation of the incoming spectrum.

\h In a sense, the two edges of a ``barrier" (undersized waveguide segment)
can be regarded as semi-mirrors of a Fabry--Perot configuration. The
consequent negative interference processes can lead themselves to
Superluminal transit times. These points have been exploited, e.g., by Japha
and Kurizki\cite{22} (who claimed the barrier transit mean-time to be Superluminal
provided that the coherence time $\tau_\crm$ of the entering field
$\psi_\inrm(t)$ is much larger than $L/c$).

\section*{2. Transients and Signals}
With regard to the same experiments, Guenter Nimtz claimed very
recently\cite{10} that those results with evanescent waves, or
``tunnelling photons", do {\em really} imply Superluminal signal and impulse
transmission.

\h A common objection consists in recalling that, on the other hand, the speed
of the precursors cannot be larger than $c$ (a fact that appears to be enough
for satisfying the requirements of the so-called, naive Einstein causality). \
Actually, every perturbation passes through a {\em transient} state before
reaching the stationary regime; this happens also when transmitting any kind
of wave. In the case of electromagnetic waves, such a transient state is
ordinarily associated with the propagation of {\em precursors}, arriving
before the principal signal. \ For instance, the existence of Sommerfeld's
and Brillouin's precursors (the so-called first and second precursors) has
been recently stressed in refs.\cite{23}, while studying the transients in
metallic waveguides.

\h To investigate the interplay between
Einstein causality and the fact that $v_\grm \gg c$ when a signal is
transported in a metallic waveguide by a carrier-wave with $\omega_\wrm <
\omega_\crm$, one has to examine {\em simultaneously} those two effects. \
To be clear about such two effects, let us recall (at the cost of repeating
ourselves) that: \ (i) in Sect.2 of ref.\cite{5} the results of computer
simulations (based on Maxwell equations only) have been presented, which
show how the first electric perturbation, reaching any point $P$, always
travels with the speed $c$ of light in vacuum, {\em independently of the
medium}; \ (ii) in Sect.3 of the same ref.\cite{5}, however, it has been
shown, by further computer simulations based on Maxwell equations only, that
the evanescent guided-waves are endowed with a Superluminal group velocity.

\h Actually, as already mentioned, the
propagation constant $\beta
(\omega )$ is imaginary for the under-cutoff frequencies, so that the
transfer function $H(\omega )$ works only as an attenuation factor for
such (evanescent) frequencies. \ However, the higher (non-evanescent)
frequencies will be phase shifted, in such a way that $\beta (\omega )$ will
tend to its free-space value $\omega /c$ for $\omega \rightarrow \infty $.
In other words, the higher spectral components travel with speed $c$; they
are the responsible both for the finite speed of the evanescent beams, and
for the appearance of the precursors.

\h At this point, one can accept ---following Nimtz--- that a signal is
really carried (not by the precursors, but) by well-defined amplitude bumps,
as in the case of information transmission by the Morse alphabet, or the
transmission of a number e.g. by a series of equal (and equally spaced)
pulses. \ In such a case, the ``signal" can travel even at infinite speed,
in the considered situations; and,
by the above-quoted computer simulations\cite{5}, it has been verified the
important fact that the {\em width} of the arriving pulses does not change
with respect to the initial ones. \ The signal, however, seems to be unable
to overcome the transients, ``slowly'' travelling with speed $c$. \
In the (theoretical) case that a pulse were constituted by under-cutoff
frequencies only, the situation could however be rather different, since
the precursors would not exist; a crucial question being: Is that
theoretical case experimentally realizable?

\h Even if the AM signal were totally constituted by under-cutoff frequencies,
when the experiment is started (e.g., by switching on the carrier wave) one
does necessarily meet a transient situation, which generates precursors. \
One might think, therefore, of arranging a setup (permanently switched on)
which produces a stable carrier-wave, so that the precursors are sent out long
before the instant at which the need arises of transmitting a signal
with Superluminal speed (without violating the naive ``Einstein causality",
as far as it requires only that the precursors do not travel at speed higher
than $c$).

\h The authors in refs.\cite{3,21,24}, do actually claim
that they can build up (smooth) signals by means of under-cutoff frequencies
only, {\em without generating further precursors\/}: in such a case one
would be in presence, then, of Superluminal information transmission. \ On
the basis of other calculations (which imply the existence also of
above-cutoff frequencies in any signal) this does not seem to be true
in practice. \ But perhaps this is not so important, in the case when a
carrier-wave is permanently switched on; in fact, due to the linearity of the
Maxwell equations, one {\em might} expect to meet solutions in which one gets both
$c$-speed precursors and faster-than-light signals: the precursors
constituting no obstacle to the passage of the Superluminal signals.

\h Such critical issues deserve further investigation. [For instance, a problem
is whether one must already know
the whole information content of the signal when {\em starting} to send it;
in such a case, it would become acceptable the mathematical trick of
representing any signal by an analytical function\cite{25}.] \ But here we want to
exploit the fact that, even if Nimtz\cite{10} is right and we are in presence
of really tachyonic motions, nevertheless nobody will be able to take
advantage of such motions for sending signals backwards in time and killing
his parents before his own birth...

\section*{3. G.Nimtz's recent claims}
Let us re-emphasize that Nimtz has very recently\cite{10} declared that the
cited results with evanescent waves do really imply Superluminal signal and
impulse transmission. \ Nimtz bases his argument on the fact that ---he
claims--- he can have recourse to a finite frequency band (entirely under
cutoff), so to have no frequencies above cutoff, and therefore no precursors
(and no front-wave) travelling at the speed of light, which could be an
obstacle to the transmission of Superluminal signals. \ An important
consequence is that Nimtz may then claim to be in presence of something
(information, and impulse) which travels faster than light.

 \h To show that he can use a finite frequency-band, Nimtz argues that his
frequency-band {\em must} be finite, since the radiated energy is of course
finite and therefore the beam cannot contain at all photons too much
energetic, i.e. with too high frequency.

\h Such a ``quantum" consideration is not as obvious as it could seem, and
should be revised and deepened since in quantum physics the Heinsenberg
uncertainty relations are known to allow, in a sense, energy violations for
sufficiently short times. \ But this is not so essential for the conclusions
in ref.\cite{10} due to the fact that, as we have already commented, the
precursors might be unable to prevent the passage of Superluminal signals in
the case when a standing wave is permanently switched on.

\h One could then tentatively {\em accept} the revolutionary claim by Nimtz.

\h However, in ref.\cite{10} it is moreover stated that one should {\em then}
accept that Einstein causality, and Special Relativity (SR), have been
violated. \ In this note we want to object to the {\em last} conclusion,
since in our opinion the existence of something travelling really
faster-than-light does not rule out the ordinary postulates of SR. In fact,
when Special Relativity (SR) is {\em not restricted} to subluminal speeds,
one ends up with an ``extended relativity" which ---on the basis of the ordinary
postulates--- can cope with actually Superluminal objects or waves without
abandoning the principle of retarded causality: Namely, without allowing one
to take advantage of those objects or waves for killing his parents before
his own birth.

\h We are going to show, in other words, that all the mentioned experimental
results do {\em not} place relativistic causality in jeopardy, even if they
refer to actual tachyonic motions. \ As far as the foundations of Extended
Relativity are concerned, we shall here confine ourselves only to quote
ref.\cite{26} and references therein; \ by contrast, we shall address our
attention to the fact that it is possible to solve the known causal paradoxes,
devised for ``faster than light" motion. \

\section*{4. Superluminal motions and relativistic causality}
Claims exist since long that all the ordinary causal paradoxes proposed
for tachyons can be solved\cite{27,28,29} (at least ``in microphysics")
on the basis of the ``switching procedure" (swp) introduced
by St\"{u}ckelberg\cite{30}, Feynman\cite{30}, Sudarshan\cite{27}, and
Recami et al.\cite{26,28}, also known as the reinterpretation principle:
a principle which in refs.\cite{26,28} has been given the status of a
fundamental postulate of special relativity, both for bradyons
[slower-than-light particles] and for tachyons. \ Schwartz,\cite{31} at last,
gave the swp a formalization in which it becomes ``automatic".

\h However, the effectiveness of the swp for the causality-problem solution
is often overlooked, or misunderstood. \
Here we want therefore to show, {\em in detail and rigorously},
how to solve the oldest ``paradox", i.e. the {\em antitelephone} one,
originally proposed by Tolman\cite{32} and then reproposed by many
authors. We shall refer to its recent formulation by Regge,\cite{33}
and spend some care in solving it, since it is the kernel of many other
paradoxes.  Let us stress that: \ (i) any careful solution of the tachyon
causal ``paradoxes" has to make recourse to explicit calculations based on
the mechanics of tachyons; \ (ii) such tachyon mechanics can be
unambiguously and uniquely derived from SR, by referring the
Superluminal ($V^2 > c^2$) objects to the class of the ordinary, subluminal
($u^2 < c^2$)  observers
{\em only} (i.e., without any need of introducing ``Superluminal
reference frames''); \ (iii) moreover, the comprehension of the whole
subject will be substantially enhanced if one refers himself to
the (subluminal, ordinary) SR based on the {\em whole} proper Lorentz group
$\Le_+ \equiv \Le^\up_+ \cup
\Le^\dow_+$, rather than on its orthochronous subgroup $\Le^\up_+$ only
[see refs.\cite{34}, and references therein]. \ At last, for a modern
approach to the classical theory of tachyons, reference can be made to
the review article\cite{26} as well as to refs.\cite{28,29}.\\

\h Before going on, let us mention the following. It is a known
fact that in the time-independent case the (relativistic,
non-quantistic) Helmholtz equation and the (non-relativistic,
quantum) Schroedinger equation are formally identical\cite{35} [in
the time-dependent case, such equations become actually different,
but nevertheless strict relations still hold between some
solutions of theirs, as it will be explicitly shown
elsewhere\cite{36}]: one important consequence of this fact being
that evanescent wave transmission simulates electron tunnelling.
On the other side, a wave-packet had been predicted\cite{37} since
long to tunnel through an (opaque) barrier with Superluminal
group-velocity. Therefore, one could expect evanescent waves too
to be endowed with Superluminal (group) speeds.\cite{38} The
abovenamed experiments\cite{1,2,3}, which seem to have actually
verified such an expectation, are the ones that most attracted the
attention of the scientific press.\cite{39} \ But they are not the
only ones which seem to indicate the existence of Superluminal
motions.\cite{40,6,7,8,9}

\section*{5. Tachyon mechanics}
In refs.\cite{41} the basic tachyon mechanics
can be found exploited for the processes: \ a) proper (or
``intrinsic") emission of a tachyon T by an ordinary body A; \ b)
``intrinsic" absorption of a tachyon T by an ordinary body A; where the term
``intrinsic" refers to the fact that those processes
(emission, absorption by A) are described {\em as they appear}
in the rest-frame of A; particle T can represent both a tachyon and an
antitachyon. Let us recall the following results only.

\h Let us first consider a tachyonic object T moving with velocity $\Vbf$ in
a reference frame $s_0$. If we pass to a second frame $s'$, endowed with
velocity $\ubf$ w.r.t. (with respect to) frame $s_0$, then the new
observer $s'$ will see ---instead of the initial tachyon T--- an
antitachyon $\ove{\rm T}$ travelling the opposite way in space (due to the
swp: cf. Appendix A), if and only if

\bb
\ubf \cdot \Vbf > c^2 \; .
\ee

\

Recall in particular that, if $\ubf \cdot \Vbf < 0$, the ``switching"
does {\em never} come into play.

\h Now, let us explore some of the unusual and unexpected
consequences of the trivial fact that in the case of tachyons it is\\

\bb
 |E| = + \sqrt{\imp^2-m^2_0} \qquad (m_0 \ \ \mbox{real}; \ {\Vbf}^2 > 1)
\; ,
\ee

\

where we chose units so that, numerically, $\: c=1$. \ We are using the
quantities ($E,\imp$), referring to the case of elementary particles; but all
our equations in the electromagnetic case can be rewritten in terms of
$\omega$ and of the wave-number $\kbf$.\ For instance (besides
$\omega^2 - \kbf^2 = 0$ for the light-like case) we shall have \ $\omega^2 -
\kbf^2 = +\Omega^2 > 0$ \ in the sublumical case, and \ $\omega^2 - \kbf^2 =
-\Omega^2 < 0$ \ in the Superluminal case: \ cf. Appendix B.

\h Let us, e.g., describe the phenomenon of ``intrinsic emission" of
a tachyon, as seen in the rest frame of the emitting body: Namely,
let us consider {\em in its rest frame} an ordinary body A, with
initial rest mass $M$, which emits a tachyon (or antitachyon) T
endowed with (real) rest mass\cite{6} $m \equiv m_0$, four-momentum $p^\mu
\equiv (E_{\rm T}, \imp)$, and velocity $\Vbf$ along the $x$-axis.
Let $M'$ be the
final rest mass of body A. The four-momentum conservation requires\\

\bb
 M = \sqrt{\imp^2-m^2} + \sqrt{\imp^2+ M'^{\, 2}} \ \ \ \ \ \ \ \
 {\rm (rest \ frame)}
\ee

that is to say [$V \equiv |\Vbf|$]:\\

\bb
 2M |\imp| = [(m^2+ \De)^2 + 4m^2 M^2]^{\frac{1}{2}} \ ; \quad V= [1+
4m^2 M^2/ (m+\De)^2]^{\frac{1}{2}} \; ,
\ee

\

where [calling  $E_{\rm T} \equiv + \sqrt{\imp^2-m^2} \,$]:\\

\bb
\De\equiv M'^{\, 2} - M^2= -m^2- 2ME_{\rm T} \; , \ \ \ \ \
{\rm (emission)}
\ee

so that\\

\bb
 - M^2 < \De \leq - |\imp|^2 \leq - m^2 \; . \ \ \ \ \ \ \ \ \
{\rm (emission)}
\ee

It is essential to notice that $\De$ is, of course, an {\em
invariant} quantity, which in a generic frame $s$ writes\\

\bb
\De= -m^2 - 2p_\mu P^\mu \; ,
\ee

\

where $P^\mu$ is the initial four-momentum of body A w.r.t. frame $s$.

\h Notice that in the generic frame $s$ the process of (intrinsic)
emission can appear either as a T emission or as a $\ove{\rm T}$ absorption
(due to a possible ``switching") by body A. \ The following theorem, however,
holds:\cite{41}\\

Theorem 1: \ $<<$ Necessary and sufficient condition for a
process to be a tachyon emission in the A rest-frame (i.e., to be
an {\em intrinsic emission\/}) is that during the process the body A
{\em lowers} its rest-mass (invariant statement!) in such a way that
$-M^2 < \De \leq -m^2$.~$>>$\\

\h Let us now describe the process of ``intrinsic absorption" of a
tachyon by body A; \ i.e., let us consider an ordinary body A to
absorb {\em in its rest frame} a tachyon (or antitachyon) T,
travelling again with speed $V$ along the $x$-direction. The
four-momentum conservation now requires\\

\bb
 M + \sqrt{\imp^2-m^2} = \sqrt{\imp^2+ M'^{\, 2}} \; , \ \ \ \ \ \ \ \
{\rm (rest \ frame)}
\ee

which corresponds to\\

\bb
\De \equiv M'^{\, 2} - M^2= -m^2+ 2 M E_{\rm T}  \; , \ \ \ \ \ \ \
{\rm (absorption)}
\ee

so that\\

\bb
 - m^2 \leq \De \leq + \infi \; .  \ \ \ \ \ \ \ \ \ \ \ \ \ \ \
{\rm (absorption)}
\ee

In a generic frame $s$, the quantity $\De$ takes the invariant form\\

\bb
 \De= -m^2+ 2 p_\mu P^\mu \; .
\ee

\

It results in the following new theorem:\\

Theorem 2: \ $<<$ Necessary and sufficient condition for a
process (observed either as the emission or as the absorption of a
tachyon T by an ordinary body A) to be a tachyon absorption in
the A-rest-frame  ---i.e., to be an {\em intrinsic absorption}---  is
that $\De \geq - m^2$.~$>>$\\

We now have to describe the {\em tachyon exchange} between two
ordinary bodies A and B. We have to consider the four-momentum
conservation at A {\em and} at B; we need to choose a (single) frame
relative to which we describe the whole interaction; let us choose the
rest-frame of A. Let us explicitly remark, {\em however}, that  ---when
bodies A and B exchange one tachyon T---  the tachyon mechanics
is such that the ``intrinsic descriptions" of the processes at A
{\em and} at B can a priori correspond to one of the following four
cases\cite{41}:

\

\setcounter{equation}{11}
\bb
\left\{\begin{array}{ll}
1) & \quad\mbox{emission---absorption} \ ,\\
\\
2) & \quad\mbox{absorption---emission} \ ,\\
\\
3) & \quad\mbox{emission---emission} \ ,\\
\\
4) & \quad\mbox{absorption---absorption} \ .
\end{array}\right.
\ee

\

Case 3) can happen, of course, only when the tachyon exchange takes
place in the receding phase (i.e., while A, B are receding from
each other); case 4) can happen, by contrast, only in the
approaching phase.

\h Let us consider here only the particular tachyon exchanges in
which we have an ``intrinsic emission" at A, and in which moreover the
velocities $\ubf$ of B and $\Vbf$ of T w.r.t. body A are such that $\ubf
\cdot \Vbf > 1$. \ Because of the last condition and the consequent
``switching" (cf. Eq.(1)), from the rest-frame of B one will therefore
observe the flight of an antitachyon $\ove{\rm T}$ {\em emitted} by B and
absorbed by A \ (a {\em necessary} condition for this to happen, let us
recall, being  that A, B  {\em recede} from each other).

\h More generally, the dynamical conditions for a tachyon to be
exchangeable between A and B can be shown to be the
following:\\

I) \ Case of ``intrinsic emission" at A:
\bb
\left\{\begin{array}{l}
 \ \mbox{if} \ \ubf \cdot \Vbf < 1 \ , \quad\mbox{then} \ \De_{\rm B} > -
m^2 \quad (\longr \mbox{intrinsic absorption at B});\\
\\
 \ \mbox{if} \ \ubf \cdot \Vbf > 1 \ , \quad\mbox{then} \ \De_{\rm B} < -
m^2 \quad (\longr \mbox{intrinsic emission at B}).\\
\end{array}\right.
\ee

II) \ Case of ``intrinsic absorption" at A:
\bb
\left\{\begin{array}{l}
 \ \mbox{if} \ \ubf \cdot \Vbf < 1 \ , \quad\mbox{then} \ \De_{\rm B} < -
m^2 \quad (\longr \mbox{intrinsic emission at B});\\
\\
 \ \mbox{if} \ \ubf \cdot \Vbf > 1 \ , \quad\mbox{then} \ \De_{\rm B} > -
m^2 \quad (\longr \mbox{intrinsic absorption at B}).\\
\end{array}\right.
\ee

\h Now, let us finally pass to examine the Tolman paradox.

\section*{6. The paradox}
\begin{figure}[!h]
\begin{center}
 \scalebox{.75}{\includegraphics{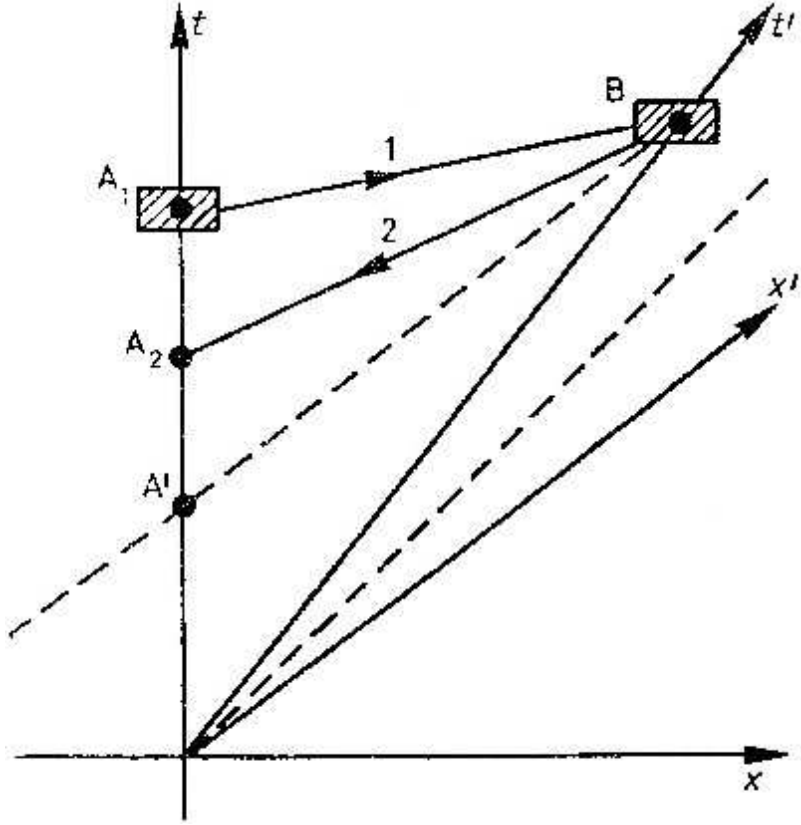}}
\end{center}
\caption{The apparent chain of the events, according to the terms of
Tolman's Paradox.}
\label{fig1}
\end{figure}

\begin{figure}[!h]
\begin{center}
 \scalebox{.75}{\includegraphics{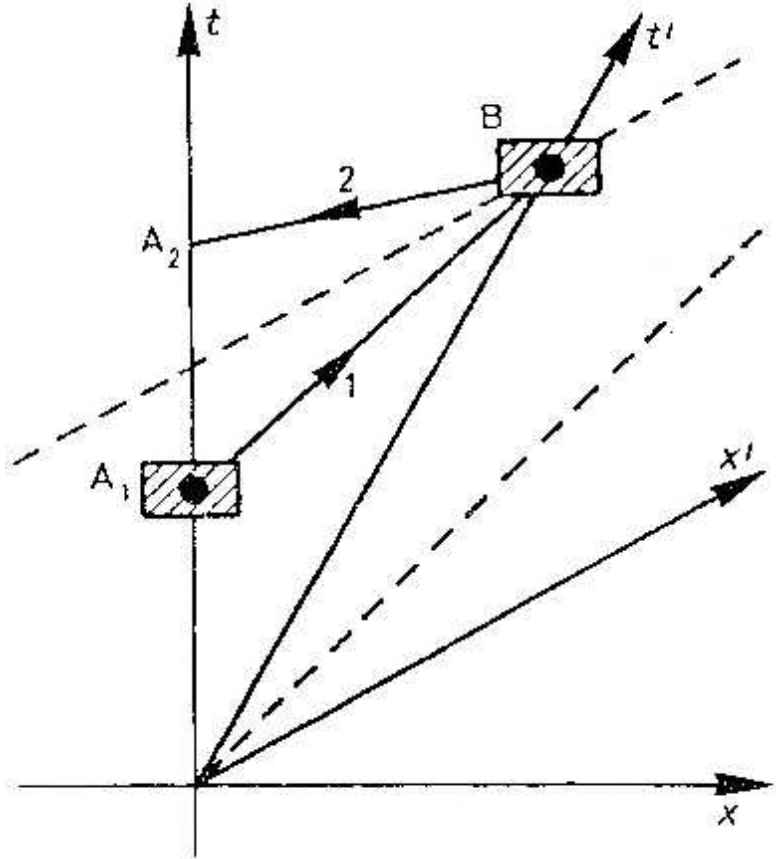}}
\end{center}
\caption{The solution of the Tolman paradox: see the text.}
\label{fig2}
\end{figure}

In Fig.\ref{fig1} and Fig.\ref{fig2}, the axes $t$ and $t'$ are
the world-lines of two devices A and B, respectively, which are
able to exchange tachyons and move with constant relative speed
$u$, [$u^2 < 1$], along the $x$-axis. According to the terms of
the paradox (Fig.1), device A sends tachyon 1 to B (in other
words, tachyon 1 is supposed to move forward in time w.r.t. device
A). The device B is constructed so as to send back tachyon 2 to A
as soon as it receives tachyon 1 from A. If B has to {\em emit}
(in its rest-frame) tachyon 2, then 2 must move forward in time
w.r.t. device B; that is to say, the world-line ${\rm BA}_2$  must
have a slope {\em lower} than the slope ${\rm BA}'$ of the
$x'$-axis (where ${\rm BA}' // x'$):
 \ this means that ${\rm A}_2$ must stay above ${\rm A}'$. If the speed of
tachyon 2 is
such that ${\rm A}_2$ falls between ${\rm A}'$ and ${\rm A}_1$, it seems
that 2 reaches A (event ${\rm A}_2$) {\em before} the emission of 1
(event ${\rm A}_1$).
This appears to realize an {\em anti-telephone}.

\subsection*{The solution}
First of all, since tachyon 2 moves
backwards in time w.r.t. body A, the event ${\rm A}_2$ will appear to
A as the emission of an antitachyon $\ove{2}$. \ The observer ``$\: t \:$"
will see his own device A (able to exchange tachyons) emit
successively towards B the antitachyon $\ove{2}$ and the tachyon 1.

\h At this point, some supporters of the paradox (overlooking tachyon
mechanics, as well as relations (12)) would say that, well, the
description put forth by the observer ``$\: t \:$" can be orthodox, but
then the device B is no longer working according to the stated programme,
because B is no longer emitting a tachyon 2 on receipt of tachyon
1. \ Such a statement would be wrong, however, since the fact that
``$\: t \:$" observes an ``intrinsic emission" at ${\rm A}_2$ {\em does not
mean} that ``$\: t' \:$" will see an ``intrinsic absorption" at B! \ On the
contrary, we are just in the case considered above, between eqs. (12)
and (13): intrinsic emission by A, at ${\rm A}_2$, with $\ubf \cdot
\Vbf_{\ove{2}} > c^2$, where $\ubf$ and $\Vbf_{\ove{2}}$ are the velocities of
B and $\ove{2}$ w.r.t. body A, respectively; so that {\em both}
A {\em and} B experience an intrinsic {\em emission} (of tachyon 2 or of
antitachyon $\ove{2}$) in their own rest frame.

\h But the tacit premises underlying the  ``paradox" (and even the very
terms in which it was formulated) were ``cheating" us
{\em ab initio}. \ In fact, Fig.1 makes it clear that, if $\ubf \cdot
\Vbf_{\ove{2}} > c^2$, then for tachyon 1 {\em a fortiori} $\ubf \cdot \Vbf_1 >
c^2$, where $\ubf$ and $\Vbf_1$ are the velocities of B and 1 w.r.t. body
A. \ Therefore, due to the previous consequences of tachyon mechanics,
observer
``$t' \,$" will see B intrinsically {\em emit} also tachyon 1 (or, rather,
antitachyon $\ove{1}$). \ In conclusion, the proposed chain of events does
{\em not} include any tachyon absorption by B (in its rest frame).

\h For body B to {\em absorb} (in its own rest frame) tachyon 1,
the world-line of 1 ought to have a slope {\em higher} than the slope
of the $x'$-axis (see Fig.2). Moreover, for body B to {\em emit}
(``intrinsically") tachyon 2, the slope of 2 should be lower
than the $x'$-axis'. In other words, when the body B, programmed to
emit 2 as soon as it receives 1, does actually do so, the event ${\rm A}_2$
does happen {\em after} ${\rm A}_1$ (cf. Fig.2), as requested by
causality.

\subsection*{The moral}
The moral of the story is twofold: \ i) one
should never {\em mix} the descriptions (of one phenomenon)
yielded by different observers; otherwise ---even in ordinary
physics--- one would  immediately meet contradictions: in Fig.1, e.g., the
motion direction of 1 is assigned by A and the motion-direction of
2 is assigned by B; this is ``illegal"; \ ii) when proposing a problem
about tachyons, one must comply\cite{27} with the rules of tachyon
mechanics\cite{38}; this is analogous to complying with the laws of
{\em ordinary} physics when formulating the text of an {\em ordinary}
problem (otherwise
the problem in itself will be ``wrong").

\h Most of the paradoxes proposed in the literature suffered the
above shortcomings.

\h Notice once more that, in the case of Fig.1, neither A nor B regard
event ${\rm A}_1$ as the cause of event ${\rm A}_2$ (or {\em vice-versa}).
In the case of Fig.2, on the other hand, both A and B consider event
${\rm A}_1$ to be the cause of event ${\rm A}_2$: but in this case
${\rm A}_1$ does
chronologically precede ${\rm A}_2$ according to both observers, in
agreement with the relativistic covariance of the law of retarded
causality.

\section*{Acknowledgements}
\h The authors gladly acknowledge stimulating discussions with
F.Bassani, G.Battiston, R.Bonifacio, M.Brambilla, R.Chiao, C.Cocca, F.Hehl,
H.E.Hern\'andez F., D.Jaroszynski, G.Kurizki, M.Mattiuzzi,
G.La Pietra, J.Le\'on, G.Nimtz, R.Petronzio, P.Pizzochero,
G.Privitera, R.Riva, L.Salvo, D.Stauffer, A.Steinberg, J.W.Swart,
C.Ussami, M.Zamboni-Rached and particularly with D.Ahluwalia and G.Degli
Antoni. \ They thank also Prof.U.Gerlach for a careful reading of
an early version of this manuscript; Prof.G.Nimtz for having made a copy of
his article available to them before publication; and the Editor and Referee
for kind interest and useful comments.\\

\

\

{\em [two Appendices follow below:]}

\newpage

{\bf APPENDIX A:} The Stueckelber--Feynman--Sudarshan--Recami
``Switching Principle" \\

\begin{figure}[!h]
\begin{center}
 \scalebox{.5}{\includegraphics{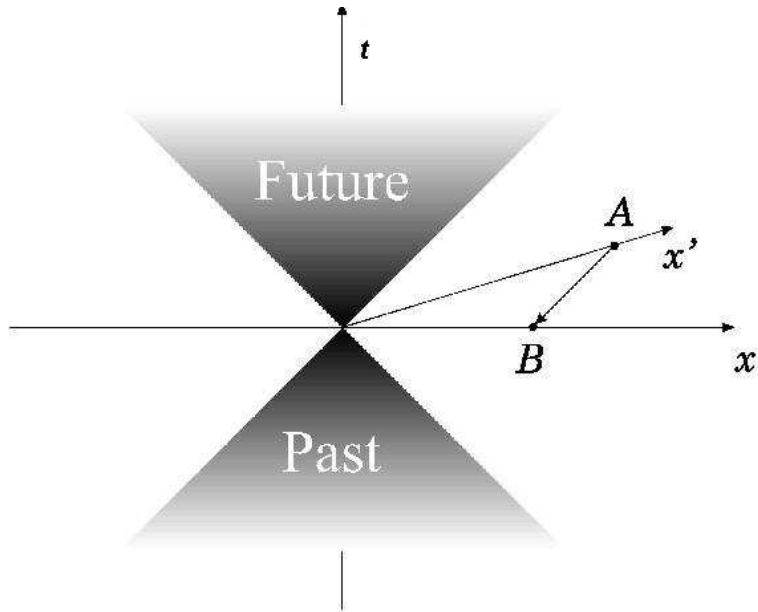}}
\end{center}
\caption{Special Relativity, when {\em not} restricted to
subluminal motions (i.e., ``Extended Relativity": see the text),
implies that any particle travelling backwards in time (from A to
B) {\em has to} carry negative energy. The simultaneous occurrence
of such two unorthodox situations will necessarily lead to an
orthodox reinterpretation, in terms of an antiparticle regularly
travelling from B to A (with positive energy and forward in time):
cf. Fig.\ref{fig5}.} \label{fig3}
\end{figure}

\h What follows refers equally well to bradyons and to tachyons.
For simplicity, then, let us fix our attention in this Appendix
only to the case of bradyons. Let us start from a positive-energy
particle P travelling forward in time. As well-known, any
orthochronows LT, \ $L^\uparrow$, \ transforms it into another
particle still endowed with positive energy and motion forward in
time. On the contrary, any antichronous (=non-orthochronous) LT, \
$L^\downarrow = - L^\uparrow$, \ will change sign --among the
others-- to the time-components of {\em all the four-vectors}
associated with P. Any $L^\downarrow$ will transform P into a
particle P' endowed in particular with negative energy {\em and}
motion backwards in time (Fig.\ref{fig3}). \ We are of course
assuming that $<<$negative-energy objects travelling forward in
time do {\em not} exist$>>$. (Elsewhere this Assumption has been
given by us the status of a fundamental postulate).

\begin{figure}[!h]
\begin{center}
 \scalebox{.5}{\includegraphics{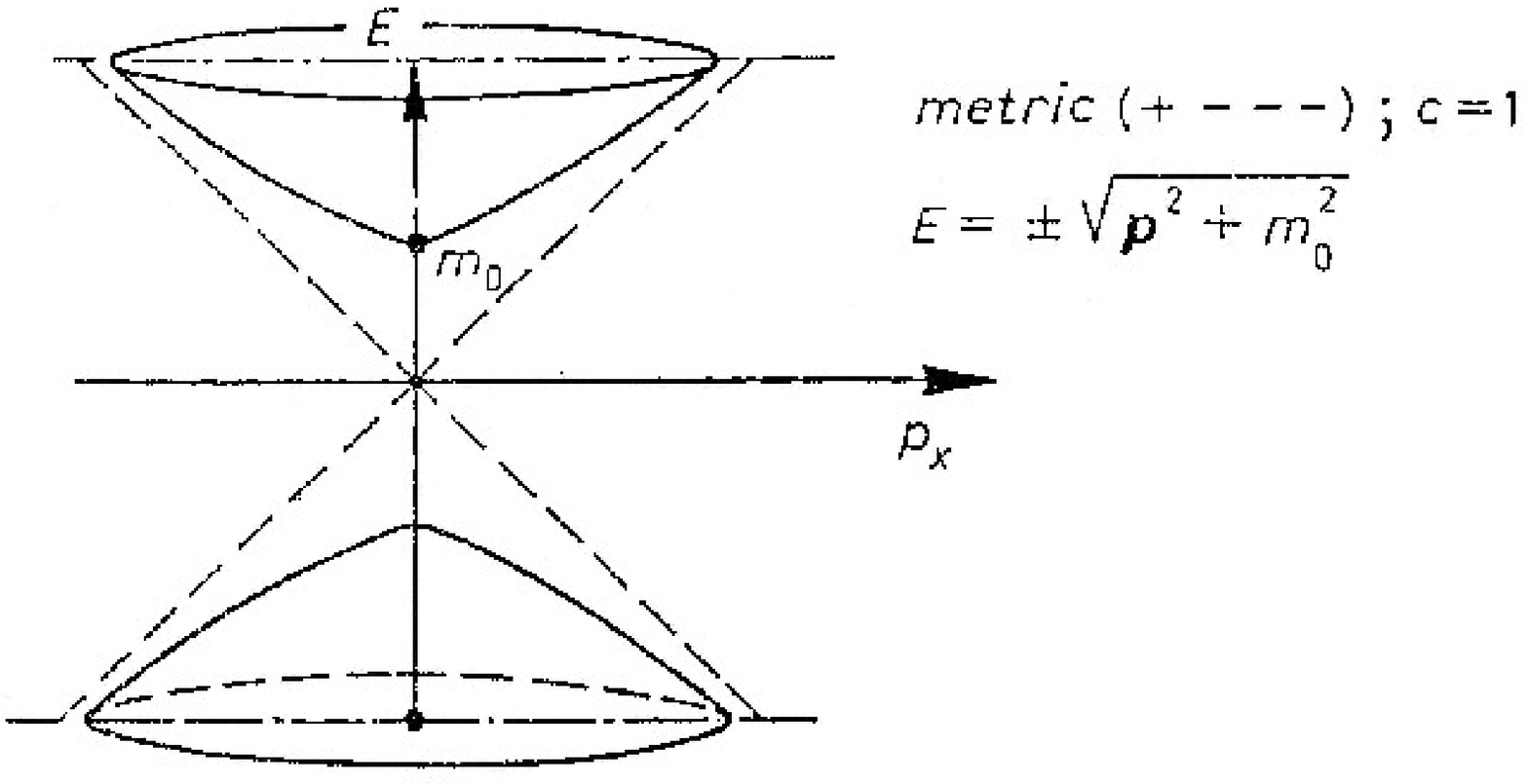}}
\end{center}
\caption{The two hyperboloid sheets, representing the kinematical states
of free particles (upper sheet) and antiparticles (lower sheet), in the
ordinary subluminal case: see the text.}
\label{fig4}
\end{figure}

\h In other words, SR together with the natural Assumption above,
{\em implies} that a particle going backwards in time
(G\"{o}del\cite{42}) (Fig.3) corresponds in the four-momentum
space, Fig.\ref{fig4}, to a particle carrying negative energy;
and, vice-versa, that changing the energy sign in the latter space
corresponds to changing the sign of time in the former (dual)
space. It is then easy to see that these two paradoxical
occurrences (``negative energy" and ``motion backwards in time")
give rise to a phenomenon  that any observer will describe in a
quite {\em orthodox} way, when they are --as they actually are--
simultaneous (Recami\cite{26,27,28,29} and refs. therein).

\

\h Notice, namely, that: \ (i) every observer (a macro-object) explores
space-time, Fig.3, in the positive $t$-direction, so that we shall
meet $B$ as the first and $A$ as the last event; \ (ii) emission
of positive quantity is equivalent to absorption of negative
quantity, as $(-) \cdot (-) = (+) \cdot (+)$; and so on.

\begin{figure}[!h]
\begin{center}
 \scalebox{.8}{\includegraphics{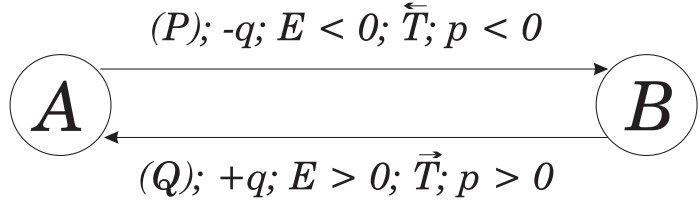}}
\end{center}
\caption{The reinterpretation (or {\em switching\/}) procedure,
mentioned in the caption of Fig.\ref{fig3}. \ The application of
such a procedure is not only possible, but compulsory: see the
text.} \label{fig5}
\end{figure}

\h Let us now suppose (Fig.\ref{fig5}) that a particle P' with
negative energy (and, e.g., charge $-e$), travelling backwards in
time, is emitted by A at time $t_1$ and absorbed by B at time $t_2
< t_1$. Then, it follows that at time $t_1$ the object A ``loses"
negative energy and negative charge, i.e. {\em gains} positive
energy and positive charge. And that at time $t_2 < t_1$ the
object B ``gains" negative energy and charge, i.e. {\em loses}
positive energy and charge. The physical phenomenon here described
is nothing but the exchange {\em from} B {\em to} A of a particle
Q with {\em positive} energy, charge $+e$, and travelling {\em
forward} in time. Notice that Q has, however, charges {\em
opposite} to P'; this means that the present ``switching
procedure" (previously called also ``RIP") effects a ``charge
conjugation" $C$, among the others. Notice also that ``charge",
here and in the following, means {\em any} additive charge; so
that our definitions of charge conjugation, etc., are more general
than the ordinary ones (see Recami and Mignani,\cite{43} hereafter
called Review I; and refs.\cite{26,28}). Incidentally, such a
switching procedure has been shown\cite{44} to be
equivalent to applying the chirality operation $\ga_5$.\\

\

\

{\em {\bf Matter and Antimatter from SR}} --  A close inspection
shows that the application of any antichronous transformation
$L^\downarrow$, together with the switching procedure, transforms
P into an object

\bb
{\rm Q} \equiv \ove{\rm P}
\ee
which is indeed the {\em antiparticle} of P. We are saying that {\em
the concept of antimatter is a purely relativistic one}, and that, on
the basis of the double sign in $[c=1]$

\bb
E = \pm \sqrt{\imp^2 + m^2_0} ,
\ee

the existence of antiparticles could have been predicted already in 1905,
exactly with the properties they actually exhibited when later
discovered, {\em provided that} recourse to the ``switching procedure" had
been made. We therefore maintain that the points of the lower
hyperboloid sheet in Fig.4 ---since they correspond not only to
negative energy but also to motion backwards in time--- represent
the kinematical states of the {\em antiparticle} $\ove{\rm P}$ {\em of}
the particle P represented by the upper hyperboloid sheet).

\h Let us stress that the switching procedure not only can, but {\em
must} be enforced, since any observer can do nothing but explore
spacetime along the positive time direction. That procedure is an improved
translation into a purely relativistic language of the
St\"{u}ckelberg--Feynman\cite{30} ``Switching principle". Together with
our  Assumption above, it can take the form of a ``Third Postulate":
$\lan\lan$Negative-energy objects travelling forward in time do {\em not}
exist; any negative-energy object P travelling backwards in time
can and must be described as its antiobject $\ove{\rm P}$ going the
opposite way {\em in space} (but endowed with positive energy and motion
forward in time)$\ran\ran$. Cf. e.g. refs.\cite{45,26,28} and references
therein.\\

{\em {\bf Concluding remarks}} --  Let us go back to Fig.3. In SR,
when based only on the two ordinary postulates, nothing prevents a
priori the event $A$ from influencing the event $B$. Just to
forbid such a possibility we introduced our Assumption together
with the Switching procedure. As a consequence, not only we
eliminate any particle-motion backwards in time, but we also
``predict" and naturally explain within SR the existence of
antimatter.

\h In the case of tachyons the Switching procedure was first applied by
Sudarshan and coworkers\cite{27};  see e.g. ref.\cite{28} and refs. therein.

\h At last, it is necessary ---however--- to observe the following: \ Whenever
it is met an object (or wavelet), ${\cal O}$, travelling at Superluminal
speed, negative contributions should be expected to the ``tunnelling"
times\cite{46,5}: and this ought not to be regarded as unphysical  or at
variance with the expectations of SR. \ In fact we have just seen above that,
whenever an ``object" ${\cal O}$ {\em overcomes}
the infinite speed\cite{26,28} with respect to a certain observer,
it will afterwards appear to the same observer as its ``{\em anti}-object"
$\ove{\cal O}$ travelling in the opposite {\em space} direction. \
For instance, when passing from the lab to a frame ${\cal F}$ moving in
the {\em same} direction as the waves (or particles) entering
the undersized waveguide (or the barrier region), the
objects $\cal O$ penetrating through that waveguide or barrier (with
almost infinite speeds\cite{5,46}) will appear in the frame ${\cal F}$ as
anti-objects $\ove{\cal O}$ crossing the waveguide or barrier {\em in the
opposite space--direction}.  In the new frame ${\cal F}$,
therefore, such anti-objects $\ove {\cal O}$ would yield a {\em negative}
contribution to the traversal time: which could even result, in total,
to be {\em negative}.\\

\

\vs{1 cm}

\

{{\large{\bf APPENDIX B:}} \ Group velocity of the evanescent waves.\\

\h Let us here define the group velocity of an electromagnetic signal in the
evanescence region, as well as in a normal (not undersized) waveguide. \
We shall follow ref.\cite{47}. \  Let us start with a normal waveguide.
The wave equation will be:
\begin{equation}
\phi(t, x, y, z) = \phi_{0}(t, x) \ \erm^{\displaystyle{i \omega t - \gamma z}} \ ,
\label{e:1}
\end{equation}
where $\gamma$ is the propagation constant inside the waveguide
\begin{equation}
\gamma^{2} \equiv k^{2}_{z} \equiv - \beta^{2} \ ,
\end{equation}
so that
\begin{equation}
\beta = i k_{z} = \pm i \sqrt{k^{2}_{\rm c} - k^{2}} = \pm \sqrt{k^{2} - k^{2}_{\rm c}}
\end{equation}
where \ $k^{2}_{\rm c} = - k^{2}_{x} - k^{2}_{y}$. \
The cutoff frequency (i.e., the lowest frequency that can propagate, in the
normal way, along a waveguide having width $a$) is given by the condition
$\beta^{2} = 0$; which yields:
\begin{equation}
\sqrt{k^{2} - k^{2}_{\rm c}} = 0 \qquad \Rightarrow \qquad k^{2} - k^{2}_{\rm c} = 0
\qquad \Rightarrow \qquad k^{2} = k^{2}_{\rm c} \ .
\label{e:2}
\end{equation}
Leu us now recall that
\[
k_{x} = - \frac{i m \pi}{a} \qquad \ {\rm and} \ \qquad k_{y} = - \frac{i n \pi}{b}
\]
and that
\begin{equation}
- k^{2}_{\rm c} = k^{2}_{x} + k^{2}_{y} \qquad \Rightarrow \qquad
k^{2}_{\rm c} = \left( \frac{m \pi}{a} \right)^{2} +
\left( \frac{n \pi}{b} \right)^{2} \ .
\label{e:3}
\end{equation}
If we adopt the propagation mode
${\rm TE}_{1 0}$, relation~(\ref{e:3}) reduces to
\[
k^{2}_{\rm c} = \left( \frac{\pi}{a} \right)^{2}
\]
which \ (since $\omega^{2} = k^{2} c^{2}$) \ yields the critical frequency:
\begin{equation}
\label{e:4}
\frac{\omega^{2}}{c^{2}} = \left( \frac{\pi}{a} \right)^{2} \qquad
\Rightarrow \qquad \omega_{\rm c} = \frac{\pi}{a} c
\end{equation}
representing the minimum frequency that a (carrier)-wave can possess in
order to be able to carry a signal, for the mentioned mode, along the
waveguide with width $a$. \
The wavepacket (group)-velocity is obtained through its dispersion relation,
which in the considered case is
\[
\beta^{2} = k^{2} - k^{2}_{\rm c}
\]
and can be rewritten as
\[
\beta^{2} = \frac{\omega^{2}}{c^{2}} - \frac{\pi^{2}}{a^{2}} \  .
\]
By definition, the group velocity is given by \
$\displaystyle{\frac{\drm \omega}{\drm \beta}}$. \ Quantity
$\beta(\omega)$ is known. One gets:
\[
\frac{\drm \beta}{\drm \omega} = \frac{\omega}{c^{2}}
\sqrt{\frac{\omega^{2}}{c^{2}} - \frac{\pi^{2}}{a^{2}}}
\]
and, by use of relation~(\ref{e:4}),
\[
\frac{\drm \beta}{\drm \omega} =
\frac{\omega}{c \sqrt{\displaystyle{\omega^{2} - \omega^{2}_{\rm c}}}} \ .
\]
Finally, by inverting, we obtain
\[
\frac{\drm \omega}{\drm \beta} = v_{\rm gr} =
c \sqrt{\displaystyle{1 - \left( \frac{\omega_{\rm c}}{\omega} \right)^{2}}}
\qquad {\rm with} \qquad \omega > \omega_{\rm c}
\]
which shows that the packet propagates in the waveguide with a
{\em subluminal} velocity.\\

{\bf Velocity of an evanescent wave} ---
If at a certain point the signal meets a ``barrier", i.e. a waveguide segment
with a smaller width $a'$, such that the lowest propagating frequency is
larger than the carrier's, the dispersion relation does then change, assuming
(as requested by Extended Relativity\cite{26}) the form:
\[
\beta^{2} = \frac{\omega^{2}}{c^{2}} + \frac{\pi^{2}}{a^{' \, 2}} \ .
\]
Following the same procedure as before, we arrive at
\[
v_{\rm gr} =
c \sqrt{\displaystyle{1 + \left( \frac{\omega_{\rm c}}{\omega} \right)^{2}}}
\qquad {\rm with} \qquad \omega < \omega_{\rm c}
\]
wherefrom it is evident that the signal is now endowed with a {\em Superluminal}
velocity.

\h Just for completeness' sake, let us add that simple simulations, based on
Maxwell equations (by using $Mathematica^{\rm TM}$), have been performed also
at Milan university\cite{47} {\em confirming these evaluations}. For instance,
a numerical experiment was performed for microwave frequencies
between 5 and 10 {\rm GHz} in a rectangular waveguide. \ Let us examine the
case with a single barrier (region II). \ In the normal barrier segments
(regions I and III, respectively), the propagating wave is described by the
equation
\[
\phi = \phi_{0} \cos \left( \frac{\pi}{a} x \right) \erm^{\displaystyle{i (\omega t - k z)}} \ ,
\]
while, in the undersized ({\em evanescent}) region, that equation becomes
\[
\phi = \phi_{0} \cos \left( \frac{\pi}{a'} x \right)
\erm^{\displaystyle{i \omega t}} \erm^{\displaystyle{-k' z}} \ .
\]
When the system is in a stationary
state, the complete set of equations is:

\[
\psi_{\Irm} = A \cos \disp{\left( \frac{\pi}{a} x \right)} \erm^{\displaystyle{i k_{1} z}} +
R \cos \disp{\left( \frac{\pi}{a} x \right)} \erm^{\displaystyle{-i k_{1} z}} \ ;
\]
\[
\psi_{\IIrm} = B \cos \disp{\left( \frac{\pi}{a'} x \right)} \erm^{\displaystyle{k_{2} z}} +
C \cos \disp{\left( \frac{\pi}{a'} x \right)} \erm^{\displaystyle{k_{2} z}} \ ;
\]
\begin{equation}
\psi_{\IIIrm} = T \cos \disp{\left( \frac{\pi}{a} x \right)} \erm^{\displaystyle{i k_{1} z}} +
E \cos \disp{\left( \frac{\pi}{a} x \right)} \erm^{\displaystyle{-i k_{1} z}} \ .
\end{equation}
If the entering waves come from $z = - \infty$ (and noone come from
$z = + \infty$), one has $A = 1$ and $E = 0$.  \ The continuity conditions,
which allow determining the coefficients $R, T, B, C$, are (for the regions
I and II)
\begin{equation}
\begin{cases}
\psi_{\rm I} (0)  = \psi_{\rm II} (0) \ ; \ \ \
\\
\displaystyle{\left. \frac{\drm \psi_{\rm I}}{\drm z} \right|_{0}}  =
 \displaystyle{\left. \frac{\drm \psi_{\rm II}}{\drm z} \right|_{0}} \ ,
\end{cases}
\end{equation}
and (for the regions II and III), quantity $L$ being the evanescence region
length,
\begin{equation}
\begin{cases}
\psi_{\rm II} (L)  =  \psi_{\rm III} (L) \ ; \ \ \
\\
\displaystyle{\left. \frac{\drm \psi_{\rm II}}{\drm z} \right|_{L}}  =
 \displaystyle{\left. \frac{\drm \psi_{\rm III}}{\drm z} \right|_{L}} \ .
\end{cases}
\end{equation}
For a carrier-wave frequency of 7 {\rm GHz}, the group velocity in the normal
and in the evanescent regions resulted to be $0.7 c$ and $1.7 c$,
respectively.\\

\vs{1 cm}

\end{document}